\documentclass[11pt,english]{article}
\usepackage{bookman}
\usepackage[T1]{fontenc}
\usepackage{geometry}
\usepackage{amsmath}
\usepackage{graphicx}

\makeatletter

\providecommand{\LyX}{L\kern-.1667em\lower.25em\hbox{Y}\kern-.125emX\@}

\usepackage{babel}
\makeatother
\begin{document}

\title{\textbf{\huge Strange Particles from NEXUS 3 }}

\author{K. Werner$^{1,\, }$%
\footnote{\noindent invited speaker, \protect \\
Strange Quark Matter Conference, Atlantic Beach, North Carolina, March
12-17, 2003%
}~, F.M. Liu$^{2,3,1,\, }$%
\footnote{\noindent Alexander von Humboldt Fellow%
}~, S. Ostapchenko$^{4,5}$, T. Pierog$^{6,1}$}

\date{$\, \, \, $}

\maketitle
\noindent \textit{\small $^{1}$ SUBATECH, Université de Nantes --
IN2P3/CNRS -- Ecole des Mines,  Nantes, France}{\small \par}

\noindent \textit{\small $^{2}$Institute of Particle Physics , Huazhong
Normal University, Wuhan, China}{\small \par}

\noindent \textit{\small $^{3}$} \emph{\small Institut fuer Theoretische
Physik, JWG Frankfurt Universitaet, Germany}{\small \par}

\noindent \textit{\small $^{4}$ Institut für Experimentelle Kernphysik,
Univ. of Karlsruhe, 76021 Karlsruhe, Germany}{\small \par}

\noindent \textit{\small $^{5}$ Moscow State University, Institute
of Nuclear Physics, Moscow, Russia  }{\small \par}

\noindent \textit{\small $^{6}$ Forschungszentrum Karlsruhe, Institut 
f\"{u}r Kernphysik, Karlsruhe, Germany}{\small \\}

\begin{abstract}
After discussing conceptual problems with the conventional string
model, we present a new approach, based on a theoretically consistent
multiple scattering formalism. First results for strange particle
production in proton-proton scattering at 158 GeV and at 200 GeV (cms)
are discussed.
\end{abstract}

\section{Problems with the String Model Approach}

\begin{figure}[htp]
\begin{center}{\Large a)}$\; $\includegraphics[  scale=0.7]{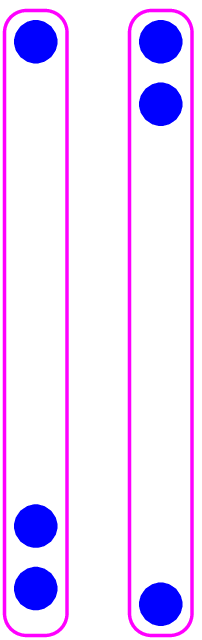}$\qquad \qquad ${\Large b)}$\: $\includegraphics[  scale=0.7]{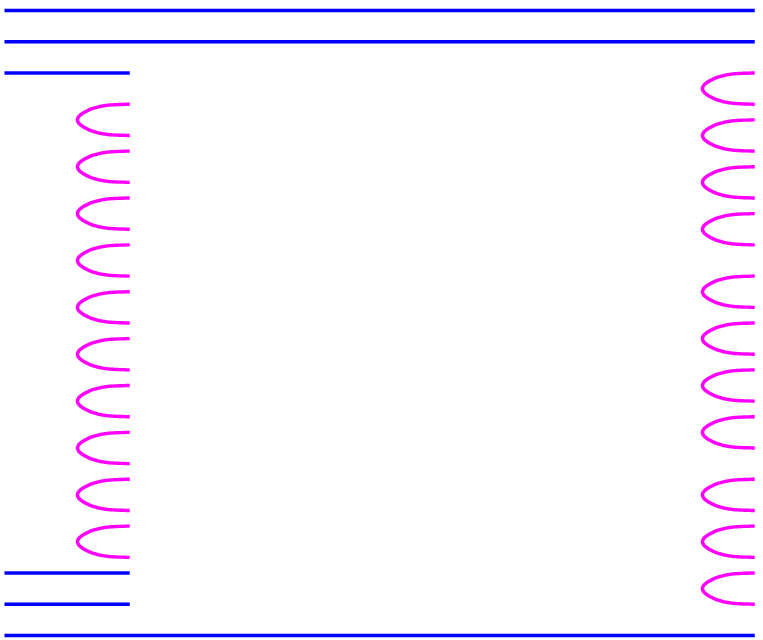}\end{center}

\caption{\label{chains}(a) A pair of strings. (b) Two chains of quark lines,
equivalent to two chains of hadrons}
\end{figure}

\begin{figure}[htp]
\begin{center}{\Large a)}$\: $\includegraphics[  scale=0.7]{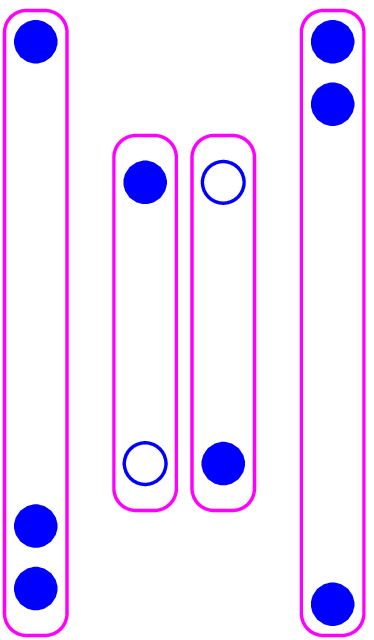}$\qquad \qquad ${\Large b)$\: $\includegraphics[  scale=0.7]{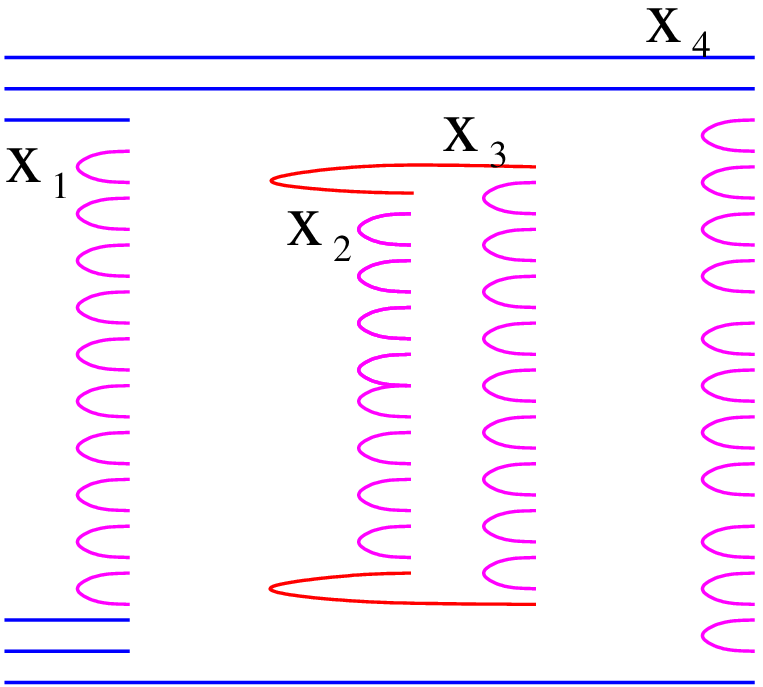}}\end{center}{\Large \par}

\caption{\label{cap:Two-pairs-of}Two pairs of strings (a) and the corresponding
chains (b)}
\end{figure}

How are string models realized? One may present the particle production
from strings via chains of quark lines \cite{ranft89} as shown in
fig. \ref{chains}. %
It turns out that the two string picture is not enough to explain
for example the large multiplicity fluctuations in proton-proton scattering
at collider energies: more strings are needed, one adds therefore
one or more pairs of quark-antiquark strings, as shown in fig. \ref{cap:Two-pairs-of}.%
 The variables $x_{i}$ refer to the longitudinal momentum fractions
given to the string ends. Energy-momentum conservation implies $\sum x_{i}=1.$

What are the probabilities for different string numbers? Here, Gribov-Regge
theory comes at help, which tells us that the probability for a configuration
with $n$ elementary interactions is proportional to $\chi ^{n}/n!$,
where $\chi $ is a function of energy and impact parameter. Here,
a dashed vertical line represents an elementary interaction (referred
to as Pomeron).

Now one identifies the elementary interactions (Pomerons) from Gribov-Regge
theory with the pairs of strings (chains) in the string model, and
one uses the above-mentioned probability for $n$ Pomerons to be the
probability for configurations with $n$ string pairs. Unfortunately
this is not at all consistent, for two reasons:

\begin{enumerate}
\item Whereas in the string picture the first and the subsequent pairs are
of different nature, in the Gribov-Regge model all the Pomerons are
identical.
\item Whereas in the string (chain) model the energy is properly shared
among the strings, in the Grivov-Regge approach does not consider
energy sharing at all (the $\chi $ is a function of the total energy
only)
\end{enumerate}
These problems have to be solved in order to make reliable predictions.

\section{NE{\LARGE X}US}

NE{\large X}US is a self-consistent multiple scattering approach to
proton-proton and nucleus-nucleus scattering at high energies. The
basic feature is the fact that several elementary interaction, referred
to as Pomerons, may happen in parallel. We use the language of Gribov-Regge
theory to calculate probabilities of collision configurations (characterized
by the number of Pomerons involved, and their energy) and the language
of strings to treat particle production. We treat both aspects, probability
calculations and particle production, in a consistent fashion: \emph{In
both cases energy sharing is considered in a rigorous way}\cite{nexo}\emph{,
and in both cases all Pomerons are identical.} This is one new feature
of our approach. Another new aspect is the necessity to introduce
remnants: The spectators of each baryon form a remnant. They will
play an important role on particle production in the fragmentation
region and at low energies ($E_{Lab}=$40-200 GeV). In the following
we discuss the details of our approach.

\begin{figure}[htp]
\begin{center}\includegraphics[  scale=0.4]{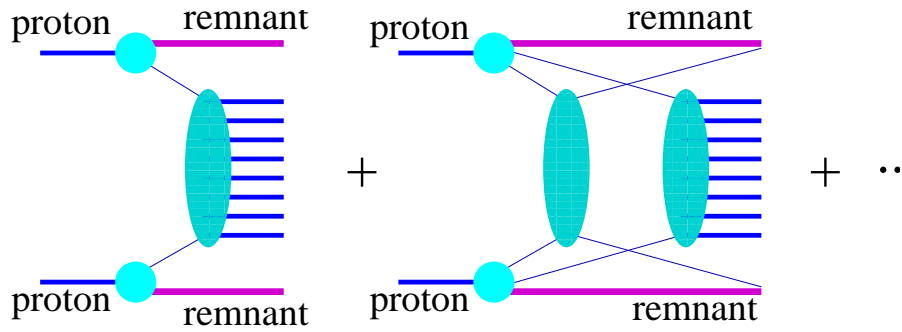}\end{center}

\caption{Inelastic scattering in pp. Partons from the projectile or the target
proton interact via elementary interactions (the corresponding produced
particles being represented by horizontal lines), leaving behind two
remnants. \label{t7}}
\end{figure}

\begin{figure}[htp]
\begin{center}(a)\includegraphics[  scale=0.4]{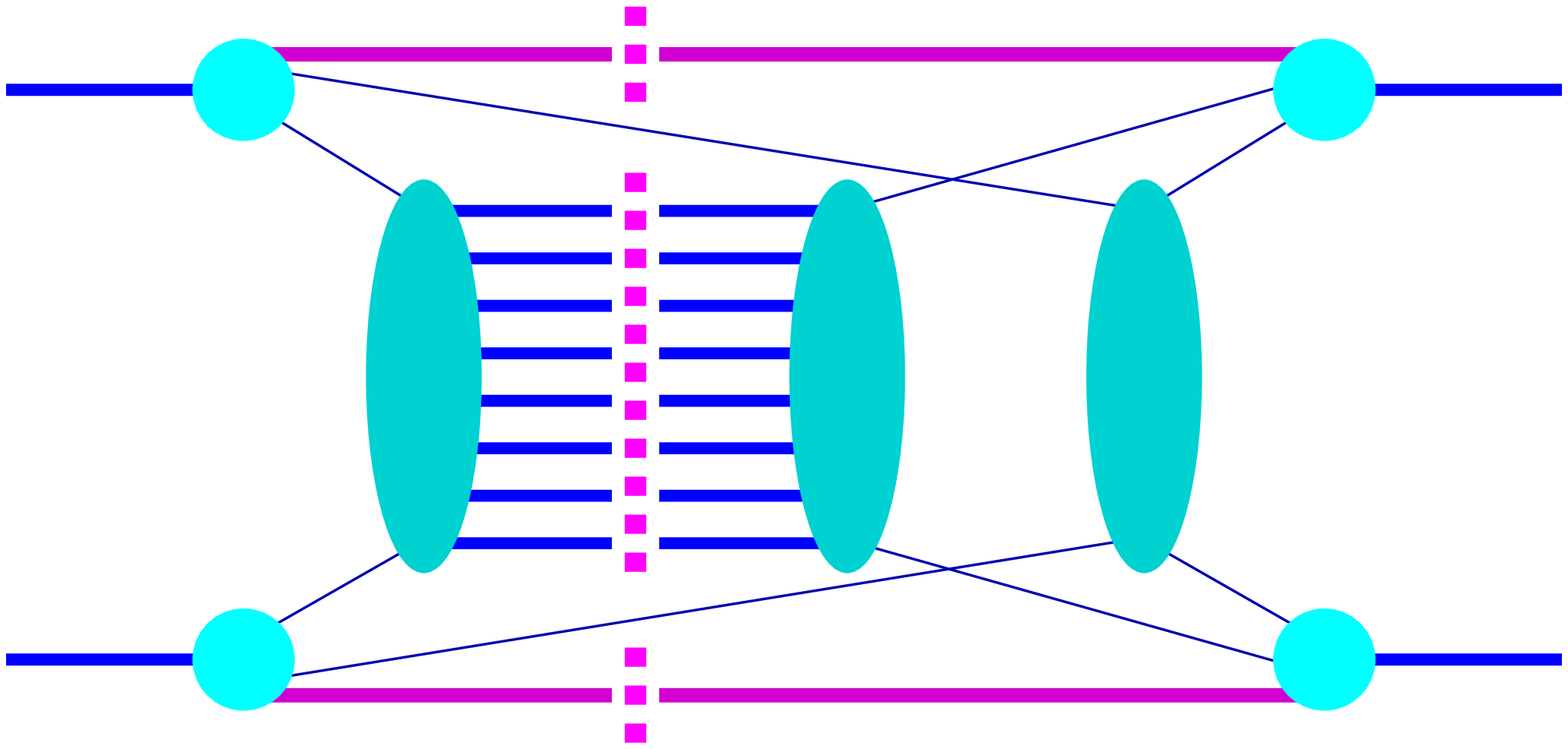}$\, \, $(b)\includegraphics[  scale=0.4]{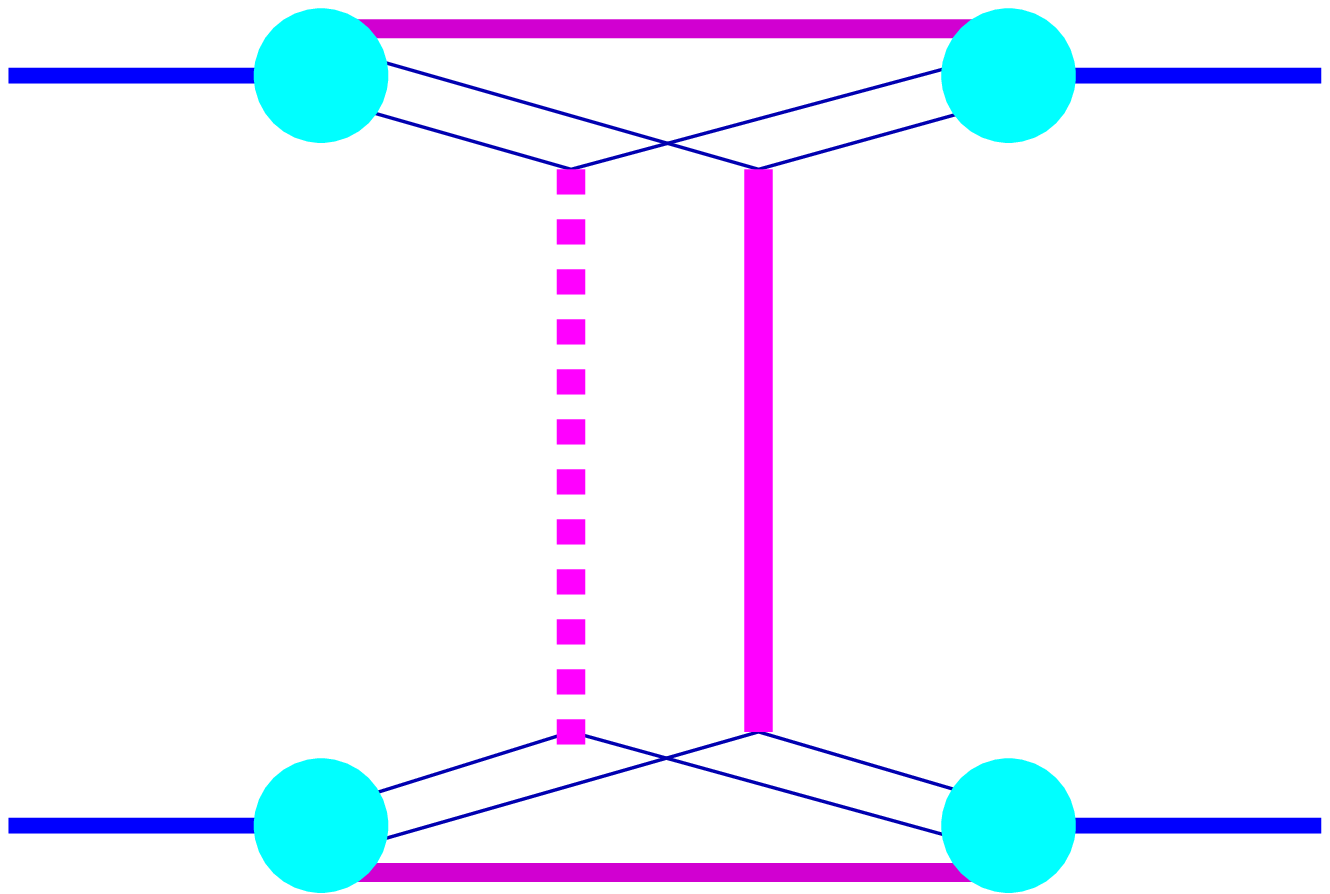}\end{center}

\caption{Inelastic scattering in pp.  a) An interference term , b) in simplified
notation.\label{t7b}}
\end{figure}

We first consider inelastic proton-proton scattering. We imagine an
arbitrary number of elementary interactions to happen in parallel,
where an interaction may be elastic or inelastic, see fig. \ref{t7}.
The inelastic amplitude is the sum of all such contributions in with
at least one inelastic elementary interaction is involved. %
To calculate cross sections, we need to square the amplitude, which
leads to many interference terms, as the one shown in fig. \ref{t7b}(a),
which represents interference between the first and the second diagram
of fig. \ref{t7}. We use the usual convention to plot an amplitude
to the left, and the complex conjugate of an amplitude to the right
of some imaginary {}``cut line'' (dashed vertical line). The left
part of the diagram is a cut elementary diagram, conveniently plotted
as a dashed line, see fig. \ref{t7b}(b). The amplitude squared is
now the sum over many such terms represented by solid and dashed lines.%

Summing appropriate classes of interference terms, we obtain probabilities
of having $m$ inelastic interactions with light cone momenta $x_{1}^{+}$..$x_{2m}^{+}$,$x_{1}^{-}$...$x_{2m}^{-}$
at a given impact parameter. Integrating over impact parameter provides
the corresponding cross section. By this we obtain a probability distribution
for the number of elementary interactions (number of Pomerons) and
the momenta of these Pomerons. 

How to form strings from Pomerons? No matter whether single-Pomeron
or multiple-Pomeron exchange happens in a proton-proton scattering,
all Pomerons are treated identically. Each Pomeron is identified with
two strings.

The string ends are quarks and antiquarks from the sea. {\small }This
differs from traditional string models, where all the string ends
are valence quarks. Due to the large number of Pomerons this is impossible
in our approach. The valence quarks stay in remnants. Being formed
from see quarks, string ends from cut Pomerons have complete flavor
symmetry and produce particles and antiparticles in equal amounts. 

\begin{figure}[htp]
\begin{center}\includegraphics[  scale=0.45]{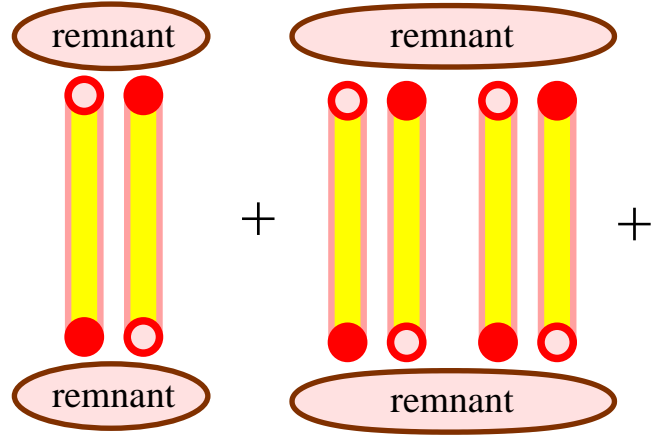}\end{center}

\caption{Remnants in single (two strings) and double scattering (four strings):
in any case, two remnants contribute.\label{remnstring2}}
\end{figure}

Remnants are new objects, compared to other string models, see fig.\ref{remnstring2}.%
 A remnant contains three valence quarks and the corresponding antiparticles
of the partons representing the string ends. We parametrize the mass
distribution of the remnant mass as $P(m^{2})\propto (m^{2})^{-\alpha }$,
where $m$ is taken within the interval $[m_{\mathrm{min}}^{2},\, x^{+}s]$,
with $s$ being the squared total energy in the center of mass system,
$m_{\mathrm{min}}$ being the minimum mass of hadrons to be made from
the remnant's quarks and antiquarks, and $x^{+}$ being the light-cone
momentum fraction of the remnant. Through fitting the data at 158
GeV, we determine the parameter $\alpha =2.25$. Remnants decay into
hadrons according to n-body phase space\cite{droplet}.

\begin{figure}[htp]
\begin{center}\includegraphics[  scale=0.8]{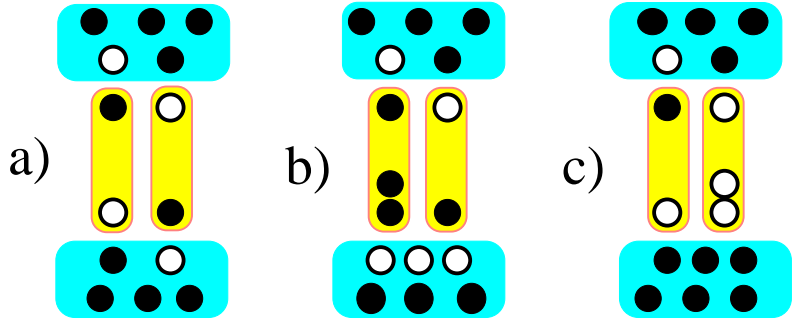}\end{center}

\caption{\label{nexus2} {\small (a) The most simple and most frequent collision
configuration has two remnants and only one cut Pomeron represented
by two $\mathrm{q}-\overline{\mathrm{q}}$ strings. (b) One of the
$\overline{\mathrm{q}}$ string-ends can be replaced by a $\mathrm{qq}$
string-end. (c) With the same probability, one of the $\mathrm{q}$
string-ends can be replaced by a $\overline{\mathrm{q}}\overline{\mathrm{q}}$
string-end. }}
\end{figure}

The most simple and most frequent collision configuration has two
remnants and only one cut Pomeron, represented by two $\mathrm{q}-\overline{\mathrm{q}}$
strings as in Fig.\ref{nexus2}(a). Besides the three valence quarks,
each remnant has in addition a quark and an antiquark to compensate
the flavor.%

We assume that an antiquark $\bar{\mathrm{q}}$ from a string end
may be replaced by a diquark $\mathrm{qq}$, with a small probability
$P_{qq}$. In this way, we get quark-diquark ($\mathrm{q-qq}$) strings
from cut Pomerons. The qqq Pomeron end (the sum of the two corresponding
string ends) has to be compensated by the three corresponding antiquarks
in the remnant, as in Fig.\ref{nexus2}(b). The (3q3$\overline{\mathrm{q}}$)
remnant decays according to phase space mainly into three mesons (3M),
and only with a very small probability into a baryon and an anti-baryon
(B+$\overline{\mathrm{B}}$). For symmetry reasons, the q string end
is replaced by an antidiquark $\overline{\mathrm{q}}\overline{\mathrm{q}}$
with the same probability $P_{\mathrm{qq}}$. This yields a $\overline{\mathrm{q}}-\overline{\mathrm{q}}\overline{\mathrm{q}}$
string and a (6q) remnant, as shown in Fig.\ref{nexus2}(c). The (6q)
remnant decays into two baryons. Since q-qq strings and $\overline{\mathrm{q}}-\overline{\mathrm{q}}\overline{\mathrm{q}}$
strings have the same probability to appear from cut Pomerons, baryons
and antibaryons are produced in the string fragmentation with the
same probability. However, from remnant decay, baryon production is
favored due to the initial valence quarks. 

With decreasing energy the relative importance for the particle production
from the strings as compared to the remnants decreases, because the
energy of the strings is lowered as well. If the mass of a string
is lower than the cutoff, it will be discarded. However, the fact
that an interaction has taken place is taken into account by the excitation
of the remnant, which follows still the above mentioned distribution.

\section{Results}

\begin{figure}[htp]
\begin{flushleft}\includegraphics[  scale=0.9]{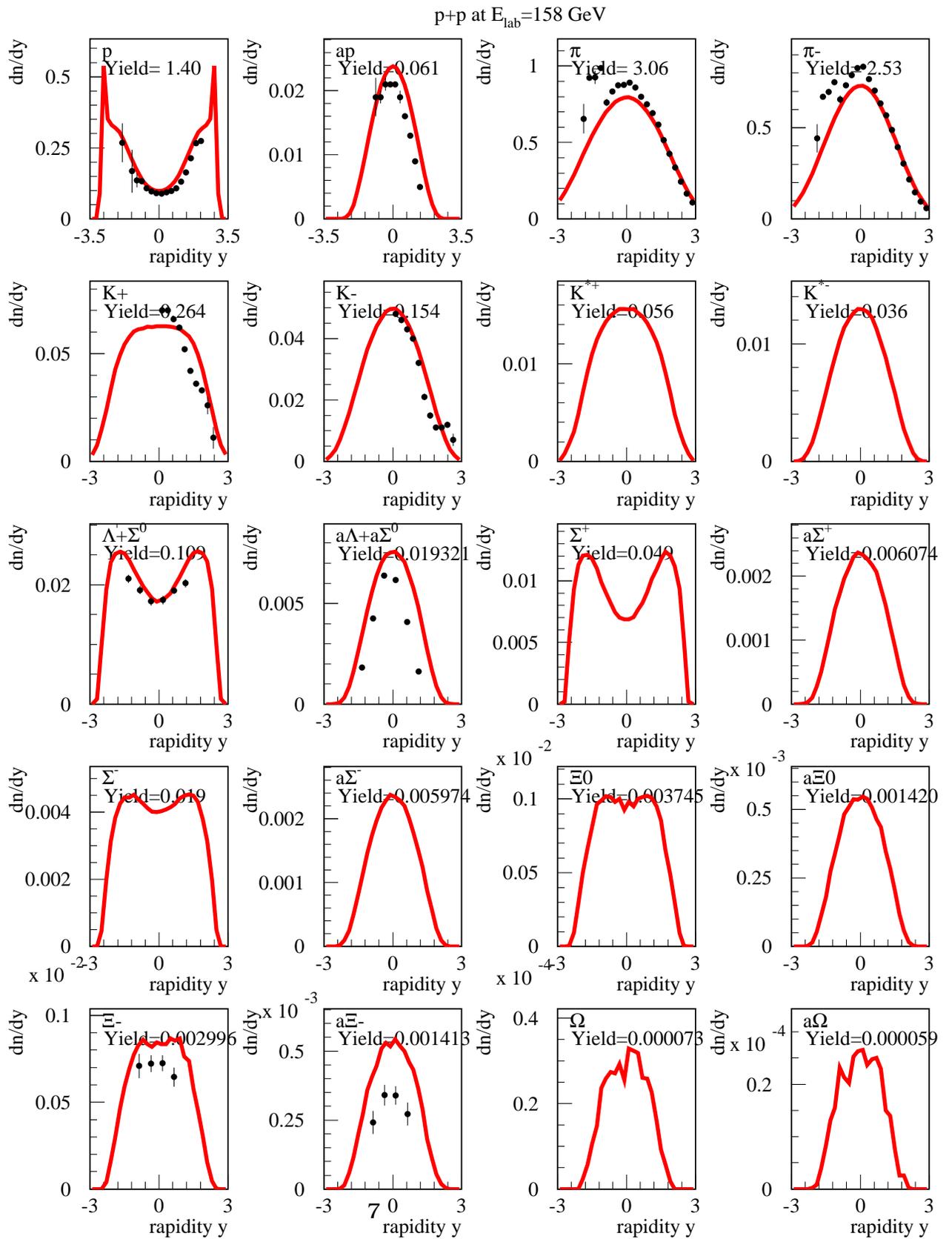} \end{flushleft}

\caption{Proton-proton at 158 GeV \label{cap:Proton-proton-at-158}}
\end{figure}

Fig. \ref{cap:Proton-proton-at-158} shows the rapidity spectra of
strange (and non-strange) hadrons for pp at 158 GeV. We included the
$\Lambda $, $\bar{\Lambda }$, $\Xi $ , $\bar{\Xi }$ spectra, which
have been published earlier \cite{nexn}. %
Where data from the NA49 collaboration are available, we have included
them in the plot\cite{dat}. The yield gives the calculated average
multiplicity of the particle species in 4$\pi $. We see that the
experimental data are reasonable described. The non strange baryons
as well as those which contain one strange quark show a double hump
structure, the others are peaked at mid rapidity. This is a consequence
of the three source structure (two remnants and Pomerons) in our approach.
The leading baryon has still the quantum number of the incoming baryon
but is moderately excited. Therefore it may disintegrate into baryons
whose quantum numbers differ not too much.

We observe in particular more $\Omega $ than $\bar{\Omega }$\cite{omeg},
as seen in experiment \cite{ppo}. This is a consequence of the modification
of NEXUS 3, explained in \cite{nexn}, as compared to the original
NEXUS 2 version \cite{nexo} (and many other string models), which
yields more $\bar{\Omega }$ than $\Omega $ due to the string topology.
We also observe more $\Omega $ than $\bar{\Omega }$ at RHIC energies,
however, the ratio is close to one, since the string contributions
dominate over remnants at high energies. We find for proton-proton
scattering at 200 GeV (cms) the ratios shown in table 1., together
with data from STAR \cite{anja}. %
\begin{table}[htp]
\begin{center}\begin{tabular}{|c|c|c|}
\hline 
&
NE{\large X}US 3.97&
STAR data\\
\hline
\hline 
$p/p$&
0.834 $\pm $ 0.003&
\\
\hline 
$\bar{\Lambda }/\Lambda $&
0.950 $\pm $ 0.005 &
0.89 $\pm $ 0.03 \\
\hline 
$\bar{\Xi }/\Xi $&
0.98 $\pm $ 0.02 &
0.97 $\pm $ 0.04 \\
\hline 
$\bar{\Omega }/\Omega $&
0.99 $\pm $ 0.08 &
0.90 $\pm $ 0.19 \\
\hline
\end{tabular}\end{center}

\caption{Antibaryon-baryon ratios.}
\end{table}

\end{document}